\documentclass[conference]{IEEEtran}
\IEEEoverridecommandlockouts

\usepackage{cite}
\usepackage{graphicx}
\usepackage{xcolor}
\usepackage[per-mode=repeated-symbol]{siunitx}
\usepackage{glossaries}
\usepackage[hidelinks, bookmarks=false]{hyperref}
\usepackage[noabbrev,capitalise]{cleveref}
\usepackage{booktabs}
\usepackage[referable]{threeparttablex}
\usepackage{tabularx}
\usepackage{array}
\usepackage[symbol]{footmisc}
\usepackage{threeparttable}
\usepackage{makecell}
\usepackage{arydshln}
\usepackage{tikz}
\usepackage{eso-pic}
\usepackage{hyphenat}
\usepackage{microtype}
\usepackage{amsmath,amssymb,amsfonts}
\usepackage{soul}
\usepackage{algorithmic}
\usepackage{enumitem}
\usepackage{multirow}
\usepackage{xspace}
\usepackage{pifont}
\usepackage{changepage}
\usepackage{multicol}
\usepackage{amssymb}
\usepackage{printlen}
\usepackage{newtxtext,newtxmath}
\usepackage{textcomp}
\def\BibTeX{{\rm B\kern-.05em{\sc i\kern-.025em b}\kern-.08em
    T\kern-.1667em\lower.7ex\hbox{E}\kern-.125emX}}

\usepackage[free-standing-units,per-mode=repeated-symbol]{siunitx}
\usepackage[allow-number-unit-breaks]{siunitx}
\DeclareSIUnit\comp{COMP}
\DeclareSIUnit\flop{FLOP}
\DeclareSIUnit\flops{FLOPS}
\DeclareSIUnit\bps{bps}
\DeclareSIUnit\Bps{Bps}
\DeclareSIUnit\gate{GE}
\DeclareSIUnit\op{OP}
\DeclareSIUnit\macu{MACU}
\DeclareSIUnit\ops{OPS}
\DeclareSIUnit\core{core}
\DeclareSIUnit\request{request}
\DeclareSIUnit\cycle{cycle}
\DeclareSIUnit\teraops{TOPS}
\DeclareSIUnit\ghz{GHz}
\DeclareSIUnit\mhz{MHz}
\DeclareSIUnit[number-unit-product = ]\percent{\%}

\definecolor{MidnightBlue}{HTML}{191970}
\definecolor{Mint}{HTML}{3EB889}
\definecolor{EnglishRed}{HTML}{A4515C}
\definecolor{SelectiveYellow}{HTML}{FFBA08}
\definecolor{CyanProcess}{HTML}{08B2E3}
\definecolor{OliveDrab7}{HTML}{4D4730}
\definecolor{Red}{HTML}{FF0000}
\colorlet{color1}{MidnightBlue}
\colorlet{color2}{Mint}
\colorlet{color3}{EnglishRed}
\colorlet{color4}{SelectiveYellow}
\colorlet{color5}{CyanProcess}
\colorlet{color6}{OliveDrab7}
\colorlet{colorAlert}{Red}
\definecolor{PulpGreen}{HTML}{168638}
\definecolor{PulpBlue}{HTML}{1269b0}
\definecolor{PulpRed}{HTML}{a8322c}
\definecolor{PulpYellow}{HTML}{f2c100}

\Crefname{equation}{Eq.}{Eqs.}
\Crefname{figure}{Fig.}{Figs.}
\Crefname{tabular}{Tab.}{Tabs.}

\newcommand\terapool[1]{\ensuremath{\text{Tera\-Pool}_{#1}}}

\makeatletter \newcommand{\AddSpaceIfAnonymous}{\@ifclasswith{acmart}{anonymous}{\vspace{10mm}}{}} \makeatother
\PackageWarning{TODO:}{#1!}

\newcommand{\circled}[1]{%
    \tikz[baseline=(char.base)]{
        \node[shape=circle,draw,inner sep=0.8pt, thick] (char) {\footnotesize\textbf{#1}};}}

\newacronym{pe}{PE}{Processing Element}
\newacronym{ai}{AI}{Artificial Intelligence}
\newacronym{gai}{GenAI}{Generative Artificial Intelligence}
\newacronym{agi}{AGI}{Artificial General Intelligence}
\newacronym{ml}{ML}{Machine Learning}
\newacronym{cpu}{CPU}{Central Processing Unit}
\newacronym{asic}{ASIC}{Application Specific Integrated Circuit}
\newacronym[longplural={Systems-on-Chip}]{soc}{SoC}{System-on-Chip}
\newacronym{fpga}{FPGA}{Field Programmable Gate Array}
\newacronym{asip}{ASIP}{Application Specific Instruction Processor}
\newacronym{gpp}{GPP}{General Purpose Processor}
\newacronym{gp}{GP}{general-purpose}
\newacronym{gpgpu}{GP-GPU}{General Purpose Graphics Processing Unit}
\newacronym{gpu}{GPU}{Graphics Processing Unit}
\newacronym{sm}{SM}{Streaming Multiprocessor}
\newacronym{cuda}{CUDA}{Compute Unified Device Architecture}
\newacronym{mpi}{MPI}{Message Passing Interface}
\newacronym{cots}{COTS}{Commercial-Off-The-Shelf}
\newacronym{soa}{SoA}{state-of-the-art}
\newacronym{roi}{ROI}{Return on Investments}
\newacronym
[
  longplural={Core Complexes}
]
{cc}{CC}{Core Complex}

\newacronym{lte}{LTE}{Long Term Evolution}
\newacronym{nr}{NR}{New Radio}
\newacronym{4g}{4G}{4th Generation}
\newacronym{5g}{5G}{5th Generation}
\newacronym{b5g}{B5G}{Beyond-5G}
\newacronym{6g}{6G}{6th Generation}
\newacronym{urll}{URLL}{Ultra-Reliable Low-Latency}
\newacronym{mmtc}{mMTC}{massive Machine Type Communications}
\newacronym{embb}{eMBB}{enhanced Mobile Broadband}
\newacronym{3gpp}{3GPP}{3rd Generation Partnership Project}
\newacronym{oran}{O-RAN}{Open-RAN}
\newacronym{ran}{RAN}{Radio Access Networks}
\newacronym{cran}{C-RAN}{Cloud Radio Access Networks}
\newacronym{gnb}{gNB}{Next Generation Node B}
\newacronym{pusch}{PUSCH}{Physical Uplink Shared Channel}
\newacronym{sdr}{SDR}{Software Defined Radio}
\newacronym{phy}{PHY}{Physical Layer}
\newacronym{cu}{CU}{Centralized Unit}
\newacronym{du}{DU}{Distributed Unit}
\newacronym{ru}{RU}{Remote Unit}
\newacronym{ue}{UE}{User Equipment}
\newacronym{ofdm}{OFDM}{Orthogonal Frequency Division Multiplexing}
\newacronym{ofdma}{OFDMA}{Orthogonal Frequency Division Multiple Access}
\newacronym{bf}{BF}{Beam Forming}
\newacronym{mimo}{MIMO}{Multiple-Input, Multiple-Output}
\newacronym{che}{CHE}{Channel Estimation}
\newacronym{dmrs}{DMRS}{Demodulation Reference Symbol}
\newacronym{tti}{TTI}{Transition Time Interval}
\newacronym{sc}{SC}{sub-carrier}

\newacronym{add}{add}{Add}
\newacronym{mul}{mul}{Multiply}
\newacronym{mac}{MAC}{Multiply\&Accumulate}
\newacronym{pmac}{p.mac}{Post-increment Multiply-accumulate}
\newacronym{axpy}{AXPY}{A Times X Plus Y}
\newacronym{dotp}{DOTP}{Dot Product}
\newacronym{sdotp}{SDOTP}{Sum Dot Product}
\newacronym{matmul}{MatMul}{Matrix Multiplication}
\newacronym{gemm}{GEMM}{General Matrix Multiplication}
\newacronym{gemv}{GEMV}{General Matrix-Vector Multiplication}
\newacronym{mvm}{MVM}{Matrix-Vector Multiplication}
\newacronym{cfft}{CFFT}{Complex Fast Fourier Transform}
\newacronym{sysinv}{SysInv}{Linear System Inversion}
\newacronym{choldec}{CholDec}{Cholesky Decomposition}
\newacronym{mmse}{MMSE}{Minimum Mean Squared Error}
\newacronym{conv2d}{Conv2D}{2D-Convolution}
\newacronym{dct}{DCT}{Direct Cosine Transform}

\newacronym{sram}{SRAM}{Static Random-Access Memory}
\newacronym{dram}{DRAM}{Dynamic Random-Access Memory}
\newacronym{spm}{SPM}{Scratchpad Memory}
\newacronym{tcdm}{TCDM}{Tightly Coupled Data Memory}
\newacronym{IDol}{I\$}{Instruction Cache}
\newacronym{dma}{DMA}{Direct Memory Access}
\newacronym{axi}{AXI}{Advanced eXtensible Interface}
\newacronym{noc}{NoC}{Nework-on-Chip}
\newacronym{csr}{CSR}{Control Status Register}
\newacronym{hbm}{HBM2E}{High Bandwidth Memory}
\newacronym{xbar}{Xbar}{Crossbar}

\newacronym{ipc}{IPC}{instructions-per-cycle}
\newacronym{wfi}{WFI}{wait-for-interrupt}
\newacronym{raw}{RAW}{read-after-write}
\newacronym{ins}{INS}{instruction}

\newacronym{fpu}{FPU}{Floating Point Unit}
\newacronym{fpss}{FP-SS}{Floating Point Sub-System}
\newacronym{ipu}{IPU}{Integer Processing Unit}
\newacronym{divsqrt}{DIVSQRT}{Division and Square-Root Unit}
\newacronym{lsu}{LSU}{Load Store Unit}
\newacronym{dsp}{DSP}{Digital Signal Processing}
\newacronym{qlr}{QLR}{Queue-Linked Register}

\newacronym{eda}{EDA}{Electronic Design Automation}
\newacronym{ge}{GE}{Gate Equivalent}
\newacronym{fo4}{FO4}{Fan-Out-of-4}
\newacronym{beol}{BEOL}{Back-End-of-Line}
\newacronym{pnr}{PnR}{Place and Route}
\newacronym{ppa}{PPA}{Power, Performance and Area}

\newacronym{numa}{NUMA}{Non-Uniform Memory Access}
\newacronym{fc}{FC}{Fully-Connected}
\newacronym{isa}{ISA}{Instruction Set Architecture}
\newacronym{simd}{SIMD}{Single Instruction Multiple Data}
\newacronym{spmd}{SPMD}{Single Program Multiple Data}
\newacronym{cdf}{CDF}{Cumulative Distribution Function}
\newacronym{api}{API}{Application Programmable Interface}
\newacronym{rtl}{RTL}{Register Transfer Level}
\newacronym{sfr}{SFR}{Synchronization Free Region}
\newacronym{dsl}{DSL}{Domain-Specific Language}
\newacronym{fifo}{FIFO}{First-In-First-Out}
\newacronym{int}{INT}{integer}
\newacronym{fp}{FP}{floating-point}

\newcommand\copyrighttext{\footnotesize \textcopyright This work has been submitted to the IEEE for possible publication. Copyright may be transferred without notice, after which this version may no longer be accessible.}
\newcommand\copyrightnotice{%
    \begin{tikzpicture}[remember picture,overlay]
        \node[anchor=south,yshift=10pt] at (current page.south) {\fbox{\parbox{\dimexpr\textwidth-\fboxsep-\fboxrule\relax}{\copyrighttext}}};
    \end{tikzpicture}%
}

\begin{document}

\bstctlcite{IEEEexample:BSTcontrol}

\title{TeraNoC: A Multi-Channel 32-bit Fine-Grained, Hybrid Mesh-Crossbar NoC for Efficient Scale-up of 1000+ Core Shared-L1-Memory Clusters
\thanks{\ifx\blind\undefined \textsuperscript{\dag}These two authors contributed equally to this work. \else Author information omitted for blind review \fi}
}
\ifx\blind\undefined
\author{%
Yichao Zhang\textsuperscript{*\dag}\quad
Zexin Fu\textsuperscript{*\dag}\quad
Tim Fischer\textsuperscript{*}\quad
Yinrong Li\textsuperscript{\S}\quad
Marco Bertuletti\textsuperscript{*}\quad
Luca Benini\textsuperscript{*\ddag}\quad
\\
{\small
 \textsuperscript{*\S}IIS, ETH Z\"{u}rich\quad%
 \textsuperscript{\ddag}DEI, University of Bologna%
}
\\
{\small\itshape%
\textsuperscript{*}\{yiczhang, zexifu, fischeti, mbertuletti, lbenini\}@iis.ee.ethz.ch\quad
\textsuperscript{\S}yinrli@student.ethz.ch %
}
}
\else
\author{
\textit{Anonymous Author(s).} \\
\textit{Submission ID: 103} \\
\\
}
\fi
\maketitle

\copyrightnotice

\begin{abstract}
A key challenge in on-chip interconnect design is to scale up bandwidth while maintaining low latency and high area efficiency.
2D-meshes scale with low wiring area and congestion overhead; however, their end-to-end latency increases with the number of hops, making them unsuitable for latency-sensitive core-to-L1-memory access.
On the other hand, crossbars offer low latency, but their routing complexity grows quadratically with the number of I/Os, requiring large physical routing resources and limiting area-efficient scalability.
This two-sided interconnect bottleneck hinders the scale-up of many-core, low-latency, tightly coupled shared-memory clusters, pushing designers toward instantiating many smaller and loosely coupled clusters, at the cost of hardware and software overheads.

We present TeraNoC, an open-source, hybrid mesh–crossbar on-chip interconnect that offers both scalability and low latency, while maintaining very low routing overhead.
The topology, built on \SI{32}{\bit} word-width multi-channel 2D-meshes and crossbars, enables the area-efficient scale-up of shared-memory clusters.
A router remapper is designed to balance traffic load across interconnect channels.
Using TeraNoC, we build a cluster with \num{1024} single-stage, single-issue cores that share a \num{4096}-banked L1 memory, implemented in \SI{12}{\nano\meter} technology.
We maximize the utilization of wiring resources by using a configurable number of read and write channels, achieving a peak bandwidth of \SI{3.74}{\tebi\byte\per\second} and a bisection bandwidth of \SI{0.47}{\tebi\byte\per\second}.
The low interconnect stalls enable high compute utilization of up to \num{0.85} IPC in compute-intensive, data-parallel key GenAI kernels.
TeraNoC only consumes \SI{7.6}{\percent} of the total cluster power in kernels dominated by crossbar accesses, and \SI{22.7}{\percent} in kernels with high 2D-mesh traffic.
Compared to a hierarchical crossbar-only cluster, TeraNoC reduces die area by \SI{37.8}{\percent} and improves area efficiency (\si{\giga\flop\per\second\per\square\milli\meter}) by up to \SI{98.7}{\percent}, while occupying only \SI{10.9}{\percent} of the logic area.
\end{abstract}

\begin{IEEEkeywords}
Many-core, network-on-chip, shared-memory
\end{IEEEkeywords}

\section{Introduction}
\label{sec:introduction}
With the rise of embodied \gls{gai}, compute architectures must support not only transformer-based computations but also edge-sensor-driven, data-parallel workloads for real-time environment interaction, all within stringent power and area constraints~\cite{Duan_2022}.
Robotics AI systems typically have full-platform power budgets below \SI{200}{\watt}~\cite{Sabo_2017}, with computing often targeting below \SI{20}{\watt}~\cite{Mohamadreza_2024}.
Furthermore, \gls{gai} scaling laws predict a $100\times$ increase in inference complexity~\cite{openai_2020}, driven by both growing model sizes and more inference steps in emerging reasoning models~\cite{wang_2025,baumann_2025}, increasing demands on memory footprint within the limited area budget of compute chips.
At the same time, model architectures evolve very rapidly, requiring flexible hardware to avoid premature obsolescence.

Scalable, programmable, but efficient many-core clusters are therefore attracting increasing attention for deployment within physical systems to enable parallel processing of diverse tasks~\cite{Rupp_2022}.
Interconnect design has become a key element for efficient cluster scaling, as its bandwidth, latency, and topology directly impact the design’s scalability and performance, in increasingly wiring-dominated scaled technologies.

\begin{figure}[htbp]
  \centering
  \vspace{-1em}
  \includegraphics[width=\linewidth]{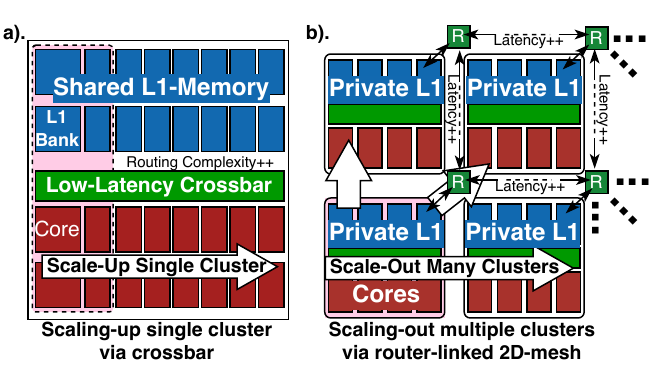}
  \vspace{-2em}
  \caption{Crossbar-based scale-up vs. 2D-mesh scale-out.}
  \label{fig:interco_scaling}
\end{figure}

Most state-of-the-art interconnects in computing clusters can be categorized into two topology templates~\cite{Biglari_2024}: \textbf{1).} low-latency, logarithmically routed \glspl{xbar}, and \textbf{2).} router-linked 2D-meshes.
\textbf{1)} offer low-latency memory access, but suffer from limited scalability because their routing complexity grows quadratically with the number of I/Os~\cite{Rahimi_2011}.
In contrast, \textbf{2)} offer better scalability thanks to their regular routing pattern, but incur latency trade-offs due to the increased number of hops.
Thus, it is commonly believed that \gls{xbar}-based interconnects are suitable only for small-scale cluster designs~\cite{Luan_2020}, where a low-complexity \gls{xbar} connects a small number of \glspl{pe} to shared memory banks (\Cref{fig:interco_scaling}a).
To meet evolving computational demands, 2D-mesh \gls{noc} is typically used to scale out many-core architectures into many loosely coupled clusters, with low intra-cluster latency, but high inter-cluster latency (\Cref{fig:interco_scaling}b).

This approach is widely adopted in modern multi-cluster designs: \emph{TensTorrent's Wormhole}~\cite{Tenstorrent_2021} features a $10 \times 8$ router-linked bidirectional 2D-torus topology.
It scales out from a \gls{xbar}-linked \num{5}-\gls{pe} cluster with \SI{1.5}{\mega\byte} shared L1 cache to a total of \num{80} clusters.
The \emph{Esperanto ET-SoC-1}~\cite{ETsoc_2022} scales out to \num{1088} cores via a 2D-mesh \gls{noc}, starting from \gls{xbar}-based clusters of \num{32} cores, with \SI{4}{\mega\byte} shared L1 cache.
\emph{HammerBlade}'s uniform \gls{noc} pattern~\cite{Hammerblade_2024} relies on a mixed 2D-mesh topology.
Each router-linked \emph{Tile} only comprises a single RISC-V core and \SI{4}{\kilo\byte} of \gls{spm}. \emph{Tiles} are linked by half-ruche (horizontal) channels to reduce cache access latency, direct mesh links connect them vertically, and skipped wormhole channels link the \emph{Cache Tiles} to HBM2 for cache refilling.
Despite this sophisticated topology, the network suffers from increasing latency as it scales out; even with support for up to \num{63} outstanding requests per \emph{Tile}, non-blocking memory access cannot be fully sustained.
Consequently, data-placement-aware programming is required to mitigate latency penalties, thereby increasing programming complexity.

Although these architectures are scalable, the loosely coupled multi-cluster organization implies major hardware and software overheads: large data structures must be split, allocated, and merged in chunks and then moved through costly, long-latency global interconnects, thereby losing energy and area efficiency in inter-cluster communication.
Recent research indicates that scaling up a single, large shared-L1-memory cluster design holds great promise for energy efficiency and ease of programming~\cite{Luan_2020}.
In \emph{NVIDIA}'s modern \gls{gpu} design, the \gls{sm} scaled-up its \gls{xbar}-connected shared memory size from \SI{192}{\kilo\byte} in \emph{A100} to \SI{256}{\kilo\byte} in \emph{H100}, while also doubling the number of \gls{fp}-\glspl{pe}~\cite{nvidia_a100_2020,nvidia_h100_2023}.
However, as \gls{xbar} routing complexity increases, area utilization suffers: the most aggressively scaled-up cluster presented in the literature, TeraPool~\cite{Zhang_2024}, features over \num{1000} RISC-V cores sharing multi-\si{\mebi\byte} memory, but requires a physical-design-aware, hierarchical multi-stage \gls{xbar} architecture that allocates up to \SI{40.7}{\percent} of the die area to routing channels.
Designing a low-latency, high-bandwidth, yet scalable and area-efficient core-to-L1-banks fine-grained interconnect is the key open challenge for efficiently scaling up shared-memory clusters.

In this paper, we tackle the bottlenecks of \gls{pe}-to-L1-memory interconnect scale-up, combining the scalability of 2D-mesh \gls{noc} with the low-latency of \glspl{xbar}.
We present TeraNoC, an open-source\footnote{\ifx\blind\undefined https://github.com/pulp-platform/TeraNoC \else Open-source information omitted for blind review. \fi}, \SI{32}{\bit} multi-channel on-chip interconnect that enables area-efficient scale-up of shared-L1-memory clusters by maximizing physical wiring utilization and minimizing area overhead.
The key contributions are:
\begin{itemize}[leftmargin=*]
    \item A hybrid Mesh–\gls{xbar} topology combining the low latency of fully combinational logarithmic \glspl{xbar} with the scalability of 2D-meshes; features low-latency, word-width, fine-grained multi-channel memory access to efficiently scale up shared-memory clusters, while fully compatible with hierarchical physical design methodologies.
    \item A router remapper that redistributes traffic load across available channels to fully exploit multi-channel bandwidth.
    \item A configurable number of read/write request channels to maximize utilization of available physical wiring resources.
    \item A physical-design-aware architecture that eases multi-channel \gls{noc} implementation; channels in the same direction can be easily bundled for routing, simplifying both floorplanning and timing closure.
\end{itemize}

We demonstrate TeraNoC within the \emph{TeraPool} cluster, the largest scaled-up shared-L1 cluster design reported in the literature~\cite{Zhang_2024}, featuring \num{1024} single-issue, single-stage cores sharing \num{4096} \SI{1}{\kibi\byte} L1 \gls{spm} banks.  
TeraNoC achieves a peak L1 bandwidth of \SI{3.74}{\tebi\byte\per\second}, while an orthogonally implemented main memory AXI interconnect by \emph{FlooNoC}~\cite{fischer_2025} reaches a peak bandwidth of \SI{9.4}{\tebi\byte\per\second} for \emph{HBM2E} access.
Compared to the hierarchical multi-stage \gls{xbar}-based \emph{TeraPool} cluster, TeraNoC reduces cluster die area by \SI{37.8}{\percent} and improves area efficiency (\si{\giga\flop\per\second\per\square\milli\meter}) up to \SI{98.7}{\percent}, while maintaining the same cluster scale.
It achieves high compute utilization (\gls{ipc} up to \num{0.85}) in key data-parallel kernels for embodied \gls{gai} workloads, demonstrating a scalable and area-efficient on-chip interconnect solution for shared-memory cluster scale-up.

In the following,~\Cref{sec:architecture} details the TeraNoC interconnect, including our proposed hierarchical design methodology and key interconnect components.  
\Cref{sec:cluster} describes the testbed cluster design and TeraNoC implementation, with \gls{ppa} results presented in~\Cref{sec:results}, followed by conclusions in~\Cref{sec:conclusion}.

\section{TeraNoC Interconnect Architecture}
\label{sec:architecture}
The key ingredients to efficiently scale up the interconnect between thousands of cores and a shared multi-thousand banked L1 memory are:
\begin{itemize}[leftmargin=*]
  \item A hierarchical design flow to achieve reasonable runtime in synthesis and physical implementation.
  \item A latency-aware topology, allowing cores' non-blocking L1 memory accessing, to keep high computing utilization.
  \item Narrow bandwidth channels for fine-grained core-to-L1-bank accesses, and large-bandwidth bundled multi-channel, routed on a regular mesh, to facilitate physical routing.
\end{itemize}
In the following subsection, we present the proposed \emph{TeraNoC} architecture, beginning with the hierarchical design overview, followed by a detailed description of each architectural element.

\begin{figure*}[ht]
  \centering
  \includegraphics[width=\linewidth]{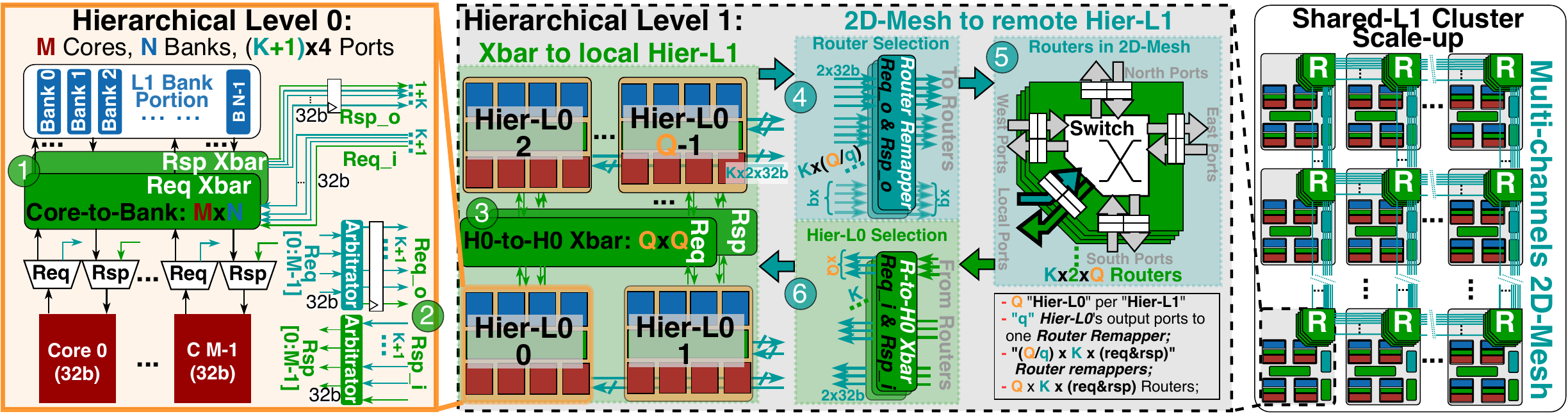}
  \vspace{-2em}
    \caption{TeraNoC hybrid mesh–crossbar topology overview in a two-level hierarchical cluster implementation. Left: base-level (\emph{Hier-L0}) crossbars connect local cores to L1 banks for low-latency access; middle: multiple \emph{Hier-L0} blocks form a higher-level (\emph{Hier-L1}) interconnected by crossbar, and routers enabling remote-\emph{Hier-L1} access; right: scalable cluster integration via a fine-grained, multi-channel 2D-mesh.}
  \vspace{-1em}
  \label{fig:interco_archi}
\end{figure*}

\subsection{TeraNoC Hierarchical Design Flow}
A hierarchical design methodology is essential to curtail front-end and back-end runtime for large-scale cluster implementations.
The cluster is partitioned hierarchically into blocks as shown in~\Cref{fig:interco_archi}: each base-level block (\emph{Hier-L0}) consists of a subset of cores and a portion of the L1 memory banks, while multiple \emph{Hier-L0} blocks are grouped into a higher-level block (\emph{Hier-L1}).
The interconnect topology at each level plays a critical role in balancing scalability, latency, and physical design feasibility.
\glspl{xbar} offer single-cycle access latency, but their routing complexity quadratically grows with the number of I/Os, limiting cluster scalability~\cite{Rahimi_2011}.
\emph{TeraPool}~\cite{Zhang_2024} adopts a multi-stage hierarchical \gls{xbar} topology to scale up clusters.
Unfortunately, the higher-level \glspl{xbar} have to reach endpoints at increasingly larger distances, hence they require a massive amount of routing area, leading to low area efficiency.
For instance, the top-level \glspl{xbar} in \emph{TeraPool} have to span an area of \SI{33.3}{\square\milli\meter}, consuming \SI{40.7}{\percent} of the cluster area.

On the other hand, the 2D-mesh \gls{noc}, known for its regular wiring and compact layout, suffers from increasing routing latency as endpoint count scales.
For example, round-trip hop counts exceed \num{60} when scaling-up to the thousand-core cluster with a flat 2D-mesh, making it difficult for cores to tolerate the latency of non-blocking accesses.
Although widening the channel width can improve throughput, core-to-memory-bank communication requires narrow, word-width channels.
This necessitates complex network interfaces between narrow-core and wide-channel protocols, and demands additional wiring resources, which would increase the cluster's floorplan area.

To address these challenges, we take inspiration from classical works in the \gls{noc} literature~\cite{Kim_2007,Grot_2009,Ludovici_2010}, and propose a hybrid Mesh–\gls{xbar} topology, combining the scalability of 2D-mesh with the low latency of \gls{xbar}.
In the base-level block (\emph{Hier-L0}), a small group of cores is connected with a portion of the L1 banks via a fully combinational \gls{xbar}, enabling low-latency, single-cycle, fine-grained bank access.
At the next level (\emph{Hier-L1}), multiple \emph{Hier-L0} blocks are connected through a second stage \gls{xbar} for intra-\emph{Hier-L1} accessing.
At the top level, \emph{Hier-L1} blocks are interconnected through a 2D-mesh \gls{noc}, which provides a regular routing pattern for the top-level connections and enables a compact layout with low routing overhead.
The \gls{noc} is designed with multiple word-width (\SI{32}{\bit}) narrow channels to support core-to-memory bank access.
The number of routers and channels ($K$) is configurable at design time to exploit available wiring resources and improve bandwidth.

Referencing~\Cref{fig:interco_archi}, when a core issues a request, arbiters at the core boundary determine whether the target bank is in the local-\emph{Hier-L0} or in a remote hierarchy block, and forward it through the \circled{1} \emph{core-to-bank} \gls{xbar} or the \circled{2} \emph{Hier-L0} block interface, respectively.
At the \emph{Hier-L0} block interface, additional arbiters select whether to forward the transaction to the next-stage \circled{3} \emph{H0-to-H0} \gls{xbar} for intra-\emph{Hier-L1} access, or \circled{4} bypass the \gls{xbar} and directly forward to the \circled{5} routers for long-distance remote \emph{Hier-L1} access via the 2D-mesh.
Once the target \emph{Hier-L1} receives the request, the \circled{6} \emph{R-to-H0} \gls{xbar} forwards the request to the destination \emph{Hier-L0} block, which then delivers it to the target memory bank through the \emph{core-to-bank} \gls{xbar}.

Interconnect dimensionality at each hierarchical level is tuned based on the following considerations:
\\\noindent\textbf{1)} The critical routing complexity (\(C_{\mathrm{Critical}}\)), is determined by the most complex \gls{xbar} in the hierarchy $i$. Designers can tune the number of \gls{xbar} inputs and outputs at each hierarchy based on wiring resources.
\begin{equation}
\begin{aligned}
& C_{\mathrm{Critical}} \simeq \max_i \left( N_{\mathrm{Inputs},i} \cdot N_{\mathrm{outputs},i} \right) \\
\end{aligned}
\label{eq:complexity}
\end{equation}
\textbf{2)} Maintain low maximum (Manhattan distance) and average (random access) round-trip latencies ($L^{\text{worst}}_{\text{2D-mesh}}$, $L^{\text{avg}}_{\text{2D-mesh}}$), which are determined by the 2D-mesh network~\cite{enright_2017}. Each hop contributes a fixed latency (\(L_{\mathrm{hop}}\)), and spill registers may be inserted to break long timing paths.
\begin{equation}
\begin{aligned}
& L^{\text{max}}_{\text{2D-mesh}} = 2\,L_{\mathrm{hop}}\,\left(2\sqrt{N_{\mathrm{Hier_{top}}}} - 1\right)\,+L_{\mathrm{SpillReg}(if\ any)} \\
& L^{\text{avg}}_{\text{2D-mesh}} \approx
\tfrac{4}{3}\,L_{\mathrm{hop}}\,\sqrt{N_{\mathrm{Hier_{top}}}}\,+L_{\mathrm{SpillReg}(if\ any)}
\end{aligned}
\label{eq:latency}
\end{equation}

\subsection{Design Elements of the TeraNoC}
This subsection details the key TeraNoC design elements, including the \gls{xbar} and mesh router design.
We also introduce a router remapper to balance 2D-mesh traffic loads and an asymmetric request channel configuration to fully exploit wiring resources.
The design parameters are summarized in~\Cref{tab:noc_parameters}.

\begin{table}[!ht]
\centering
\caption{TeraNoC Parameter Descriptions}
\begin{tabularx}{\linewidth}{rX}
\toprule
\textbf{Symbol} & \textbf{Description}                                                          \\\midrule
$Q$             & Number of \emph{Hier-L0} blocks within each \emph{Hier-L1} block              \\
$M$             & Number of cores per \emph{Hier-L0} block                                      \\
$N$             & Number of L1 memory banks per \emph{Hier-L0} block                            \\
$K$             & Number of routers per \emph{Hier-L0} for remote \emph{Hier-L1} accessing      \\
$\times 2$      & Indicates req\&rsp channels where shown in the figure                         \\
$q$             & Number of \emph{Hier-L0} ports allocated to one router remapper               \\\bottomrule
\end{tabularx}
\label{tab:noc_parameters}
\end{table}

\subsubsection{Logarithmic Crossbar}
\label{sec:crossbar_archi}
Low latency is achieved by a fully combinational, fully connected \gls{xbar} with a logarithmically staged interconnect topology~\cite{Luan_2020}.
It employs a multiplexer tree for routing and a demultiplexer for arbitration with combinational control logic~\cite{Rahimi_2011}.
The design supports fine-grained address interleaving and enables single-cycle access between cores and memory banks.
A valid–ready handshake network protocol is used:
forward requests carry metadata, including address and control signals, along with write data.
Backward responses include the initiator's address, read data, and acknowledgment.
Arbitration conflicts on the same switch are resolved using a round-robin strategy.

\subsubsection{Word-Width Fine-Grained Router}
\label{sec:router}
Communication between remote \emph{Hier-L1} blocks is enabled through the 2D-mesh \gls{noc}.
Each \emph{Hier-L0} block is equipped with $K \times 2$ ports (for requests and responses) that connect to $K \times 2$ routers, where $K$ is configurable at design time.
We build upon the FlooNoC architecture~\cite{fischer_2025}, adapting it to better suit the characteristics of L1 traffic.
In the case of TeraNoC, the core-level protocol is single-issue, word-width read and write requests and responses.
This simplicity allowed us to directly expose the core-level protocol to the router, without a complex network interface to bridge between core-level and link-level protocols.
The essential information for \gls{noc} routing, such as source and destination addresses, is already embedded in each request.
Consequently, the only required transformation involves extracting \emph{header} and \emph{payload} fields, which are used internally by the router.

The router employs dimension-ordered \emph{XY}-routing, $5\times5$ ports, and input and output \gls{fifo} buffers.
While \emph{FlooNoC} employed wide \SI{512}{\bit} links optimized for burst-based, high-bandwidth L2 traffic, TeraNoC instead utilizes narrow, fine-grained \SI{32}{\bit} links, where each router can process one core request per cycle.
To mitigate the limited bandwidth inherently associated with these narrow links, TeraNoC introduces a large number of physical channels.
First, both request and response channels are replicated $K$ times to further boost throughput.
Second, request and response paths are fully separated ($K\times2$), which prevents message-level deadlocks.

\subsubsection{Router Remapper}
Although multiple word-width \gls{noc} channels with a core-level protocol enable fine-grained L1 memory access, statically connecting each \emph{Hier-L0} block’s $K \times 2$ ports to a fixed subset of routers can cause traffic imbalance.
For example, when different \emph{Hier-L0} blocks within the same \emph{Hier-L1} access distinct target L1 spaces, their assigned routers forward transactions exclusively in their corresponding directions based on the \emph{XY}-routing protocol, while the channels in the other three directions remain idle. 
If one router becomes congested in its active direction, although available bandwidth in the same direction from other \emph{Hier-L0} blocks' routers remains, it cannot be shared due to the fixed connection assignment, leading to inefficient bandwidth utilization.

\begin{figure}[htbp]
  \centering
  \includegraphics[width=\linewidth]{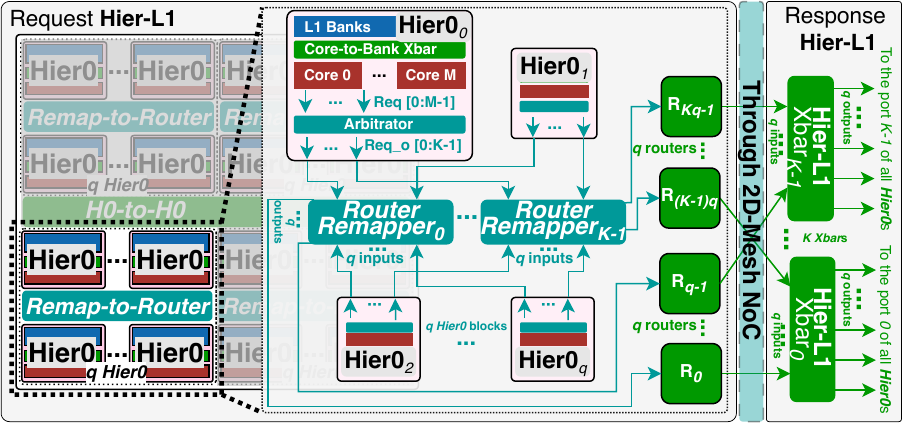}
  \vspace{-2em}
  \caption{Router remapper connections between $q$ Hier-L0 blocks' request/response ports and $q \times K$ request/response routers.}
  \vspace{-1em}
  \label{fig:router_remapper}
\end{figure}

To address this issue, in principle, all $Q \times K \times 2$ ports from the $Q$ \emph{Hier-L0} blocks within the same \emph{Hier-L1} block could be remapped to their routers using a large \gls{xbar}-based remapping mechanism.
However, the \gls{xbar} routing complexity increases quadratically with the number of inputs and outputs, making physical implementation impractical.
We decompose this remapping idea into multiple, lightweight, small-scale \gls{xbar}-based \emph{Router Remappers}, each remapping the connection between a number ($q$) of \emph{Hier-L0} ports and their routers.
\Cref{fig:router_remapper} illustrates our remapping scheme.
Each request/response remapper connects one of the $K$ ports from each of the $q$ \emph{Hier-L0} blocks, and collectively, the $K$ remappers map these $q \times K$ ports to $q \times K$ routers, enabling balanced utilization across channels.
The remapper’s control logic is implemented using a shift register initialized with a seed value to generate a pseudorandom mapping pattern, redistributing the traffic load across the sending ports to the routers.
Moreover, we observe that cores within the same \emph{Hier-L0} block typically exhibit similar traffic directions due to spatial locality in parallel computing.
Bandwidth utilization can be further improved by redistributing traffic across spatially distant \emph{Hier-L0} blocks, for example, by applying a stride-based offset on \emph{Hier-L0} IDs to remap blocks.

\subsubsection{Asymmetric Request Channels}
For most data-parallel kernels, memory access patterns exhibit a significantly higher volume of load requests than store requests.
For example, in \gls{matmul} and \gls{conv2d}, the store-to-load request ratios are only \num{0.016} and \num{0.056} per \gls{pe}.
Even in more memory-intensive kernels such as \gls{axpy} (\num{0.5}) and \gls{dotp} (\num{0.33}), load requests still dominate.

As discussed in~\Cref{sec:router}, each request includes both read and write channels, with a \emph{payload} field containing the data.
However, for load requests, which are far more frequent than store requests, the payload field is unused, yet the physical resources (wiring and buffers) for this "wider" channel are still required, losing routing utilization.
To address this inefficiency, \emph{TeraNoC} introduces two types of request channels: \emph{read-write} and \emph{read-only}.
The read-only channels omit the data payload field, making them physically "narrower" and reducing wiring complexity and buffer usage.
Furthermore, routing channels in the same direction can be easily bundled, streamlining wire routing in physical design.

\section{Large Scale-up Cluster Implementation}
\label{sec:cluster}
We compare our new design with \emph{TeraPool}, the largest many-core shared-L1-memory cluster reported in the literature~\cite{Zhang_2024}.
Each \gls{pe} is a single-issue, single-stage, \num{32}-bit \emph{RV32IA Snitch} core, extended with an \gls{ipu} (int32/16b) and an \gls{fpss} (fp32/16b).
The core’s \gls{lsu} is designed with an outstanding transaction table (\num{8} entries by default) to tolerate memory access latency.
In this section, we introduce the hierarchical \gls{xbar}-based TeraPool design as the baseline implementation, followed by the TeraNoC-based design constructed at the same cluster scale.

\subsection{Baseline: Hierarchical-Crossbar-Based Interconnect}
\emph{TeraPool} features \num{1024} \emph{Snitch} cores sharing \SI{4}{\mebi\byte} (\num{4096} banks) of L1 \gls{spm} through a multi-stage fully-connected logarithmic \glspl{xbar}, organized into three hierarchy levels.
In the base \emph{Tile} hierarchy (\emph{Hier-L0}), \gls{xbar} connects the \num{8} cores to \num{32} \SI{1}{\kibi\byte} L1 \gls{spm} banks.
Eight \emph{Tiles} are grouped and interconnected by \glspl{xbar} to form a \emph{Subgroup} (\emph{Hier-L1}), and four \emph{Subgroups} are interconnected into a \emph{Group} (\emph{Hier-L2}), with a total of four \emph{Groups} in the cluster.
This multi-stage \gls{xbar} design incorporates spill registers at hierarchical boundaries to break long timing paths, resulting in \gls{numa} latencies ranging from \num{1} to \num{9}/\num{11} cycles.
\SI{32}{\bit} data interleaved across all \gls{spm} banks to reduce bank conflicts.
However, since a \emph{Tile}'s remote requests are routed based on the target hierarchy, conflicts can occur at the hierarchy boundary arbitration ports when different cores within the same \emph{Tile} access banks in the same target hierarchy.
Furthermore, high-complexity \glspl{xbar} require a large physical routing area, occupying approximately \SI{40}{\percent} of the total die area and thereby limiting overall area efficiency.

\subsection{Our Solution: TeraNoC-Based Interconnect}
We implement TeraNoC at the same scale as the \emph{TeraPool} cluster, using the same \glspl{pe} and \gls{spm} banks organized in a two-level hierarchy.
In the \emph{Tile} design (\emph{Hier-L0}), we employ a fully connected logarithmic \gls{xbar} to connect \num{4} \emph{Snitch} cores with \num{16} \gls{spm} banks, achieving single-cycle latency for local \emph{Tile} memory accesses.
Sixteen \emph{Tiles} ($Q=16$) form a \emph{Group} (\emph{Hier-L1}), and the cluster comprises \num{16} \emph{Groups} arranged in a $4\times4$ 2D-mesh topology.

As described in~\Cref{fig:interco_archi}, each \emph{Tile} is equipped with $1+K$ remote memory access request and response ports for both incoming and outgoing, enabling interconnection with other \emph{Tiles}.
One port connects locally to other \emph{Tiles} within the same \emph{Group} via a \num{16}$\times$\num{16} logarithmic \gls{xbar}, while the remaining $K$ ports connect to $K$ TeraNoC routers for inter-\emph{Group} accessing.
The $K$ ($K \leq$ No.\glspl{pe}) is hardware-configurable and determined by the available physical routing resources in the target implementation.
In our testbed, each \emph{Tile} is connected with $K=2$ routers, each equipped with two-depth \gls{fifo} buffers for every routing direction.
To maximize bandwidth under limited routing resources, we configure two routers' request channels as one narrow \emph{read-only} and one wide \emph{read-write} channel.
To improve routing channel utilization between routers, we implement one router remapper per $q=4$ \emph{Tiles}.
On the receiving side, each \emph{Group} includes $K=2$ additional \num{16}$\times$\num{16} logarithmic \glspl{xbar}, which route incoming requests\&responses to their target \emph{Tile}.
The receiving \emph{Tile}'s remote \gls{xbar} then selects the destination \gls{spm} bank or core.
Orthogonally, each \emph{Group} integrates a \SI{512}{\bit} AXI master port, routed through the 2D mesh topology via FlooNoC routers~\cite{fischer_2025} to the \emph{HBM2E} main memory, supporting \gls{IDol} refills and \gls{dma}-managed data transfers between the L1 banks and main memory.

\section{Results}
\label{sec:results}
In the following, we analyze the TeraNoC \gls{pe}-to-L1-banks interconnect performance, physical design \gls{ppa}, and the performance of key data-parallel embodied \gls{gai} kernels.

\subsection{TeraNoC Interconnect Analysis}
\label{sec:interco_analysis}
\subsubsection{Memory Accessing Latency}
Since the TeraNoC-based testbed cluster implements a hierarchical architecture with \gls{numa} latency, we first analyze latency at each hierarchy level, followed by an overview of the entire cluster.
\begin{itemize}[leftmargin=*]
    \item \emph{Intra-Hier0}: the fully combinational logarithmic \gls{xbar} described in~\Cref{sec:crossbar_archi} is used to build the \gls{pe}-to-L1-\gls{spm} interconnect within each local \emph{Tile} (Hier-L0). Combined with the single-stage latency-tolerant \emph{Snitch} core, the interconnect guarantees a single-cycle round-trip \gls{pe}-to-L1-\gls{spm} latency, providing the fastest possible access.
    \item \emph{Intra-Hier1}: for accesses to L1 banks in other \emph{Tiles} within the same \emph{Group} (Hier-L1), requests and responses are routed through $16 \times 16$ logarithmic \glspl{xbar}. Spill registers are inserted at the outgoing boundary of each \emph{Tile} to cut the long-distance critical path, resulting in a round-trip latency of \num{3} cycles.
    \item \emph{Inter-Hier1}: for long-distance accesses to remote \emph{Groups}, requests and responses traverse the 2D-meshes through routers. The round-trip latency of the meshes is calculated using~\Cref{eq:latency}. In our $4 \times 4$ 2D-mesh topology, the per-hop latency is configured at $L_{\mathrm{hop}} = 2$ cycles. The resulting round-trip latencies (including mesh and \glspl{xbar}) are \num{7} cycles for accesses to neighboring \emph{Groups} (\num{1}-hop), \num{31} cycles to the farthest \emph{Groups} (\num{7}-hop), and \num{13.7} cycles on average.
\end{itemize}
Compared to the common approach of directly scaling up \emph{Tiles} using a 2D-mesh, TeraNoC’s hybrid interconnect achieves substantially lower latency.
For instance, in a flat $16 \times 16$ \emph{Tile} mesh, the maximum and average zero-load latencies increase to \num{127} ($4.1\times$) and \num{45.7} ($3.3\times$) cycles, respectively.
Compared to \emph{TeraPool}'s \num{1}–\num{9}/\num{11} cycles of \gls{numa} latency, TeraNoC maintains similarly low-latency access, ranging from \num{1}-\num{3} cycles within local-\emph{Group} accessing through \glspl{xbar}.
For remote-\emph{Group} accesses, compared to the \emph{TeraPool}'s multi-stage \glspl{xbar} latency of \num{5} cycles (nearest) to \num{9}/\num{11} cycles (farthest), TeraNoC achieves \num{7}-cycle latency between adjacent \emph{Groups}, while maintaining an average latency of \num{13.7} cycles.

\subsubsection{Bandwidth Analysis}
TeraNoC achieves a peak \gls{pe}-to-L1-bank bandwidth of \SI{4}{\kibi\byte\per\cycle} and a bisection bandwidth of \SI{0.5}{\kibi\byte\per\cycle} across 2D-mesh, enabling high-throughput data movement for high memory traffic tasks.
This high-throughput mesh provides \num{32} parallel word-width data response channels per direction on each router link.
In total, the $4 \times 4$ 2D-mesh topology contains \num{1536} unidirectional data response channels.
Each \emph{Tile} connects to one \emph{read-write} router and one \emph{read-only} router for remote \emph{Group} access, sustaining a bandwidth of \SI{0.5}{req\per\core\per\cycle} for read and \SI{0.25}{req\per\core\per\cycle} for write, while supporting a data response bandwidth of \SI{2}{\byte\per\core\per\cycle}.
For local accesses within the same \emph{Tile}, the bandwidth increases to \SI{1}{req\per\core\per\cycle} and \SI{4}{\byte\per\core\per\cycle} data response.
For targeting other \emph{Tiles} within the same \emph{Group}, a shared \emph{Tile} port to local-\emph{Group} \gls{xbar} provides \SI{0.25}{req\per\core\per\cycle} and \SI{1}{\byte\per\core\per\cycle} data response bandwidth.

\subsubsection{Router-Remapper Enhanced Network Utilization}
To clearly demonstrate the network utilization improvements enabled by our router remapper design, we define \emph{\gls{noc} congestion} (ChannelStalls/Cycle) as the ratio of stall cycles to total valid request cycles, capturing how often requests experience backpressure due to channel or router contention.

\begin{figure}[htbp]
  \centering
  \includegraphics[width=\linewidth]{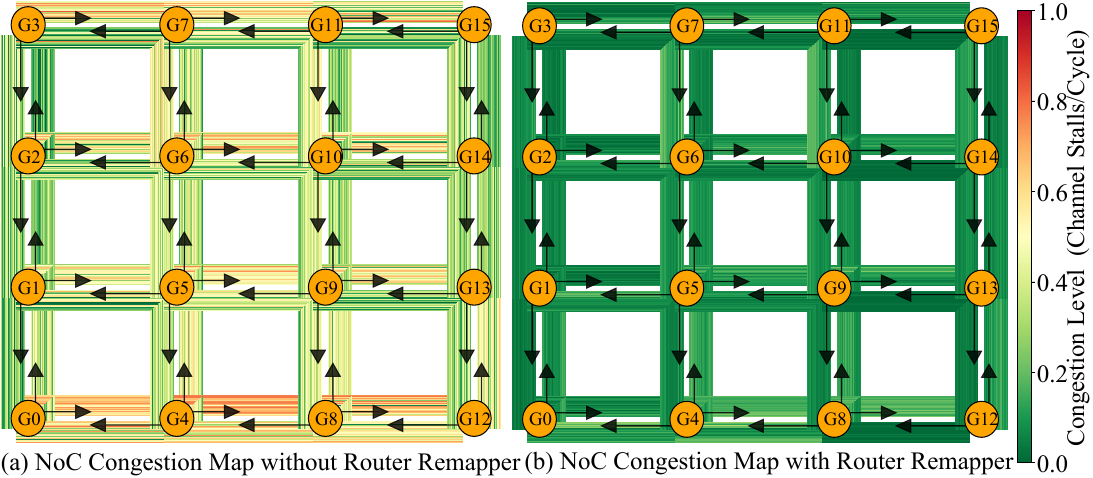}
  \vspace{-2em}
  \caption{Network utilization improvement with router remapper.}
  \label{fig:remapper_utilization}
\end{figure}

We profile \emph{\gls{noc} congestion} with the \gls{matmul}(f32) kernel, a key data-parallel workload in which each \gls{pe} aggressively fetches matrix data globally across all L1 banks.
We capture a \num{3000}-cycle trace of the inter-\emph{Group} mesh \gls{noc} traffic during the kernel’s inner loop and use it to generate the congestion heatmaps shown in~\Cref{fig:remapper_utilization}. In these visualizations, higher congestion level values indicate channels that experienced more frequent stalls during the trace window.
As shown in~\Cref{fig:remapper_utilization}(a), without the router remapper, congestion is unevenly distributed across channels, with an average congestion ratio of \SI{0.40}{ChannelStalls/Cycle} and a peak of \SI{0.83}{ChannelStalls/Cycle}.
With the remapper enabled, as shown in~\Cref{fig:remapper_utilization}(b), traffic becomes more evenly distributed, reducing the average congestion by \SI{80}{\percent} to \SI{0.08}{ChannelStalls/Cycle}, and lowering the peak by \SI{63}{\percent} to \SI{0.31}{ChannelStalls/Cycle}.
Correspondingly, the observed global L1 memory access bandwidth improves by $2.7\times$, from \SI{405.3}{\gibi\byte\per\second} to \SI{1081.4}{\gibi\byte\per\second}.
These results demonstrate the effectiveness of our lightweight router remapping mechanism in improving bandwidth utilization, mitigating localized hotspots, and enabling higher sustained throughput under high-intensity access patterns.

\begin{figure*}[ht]
  \centering
  \includegraphics[width=\linewidth]{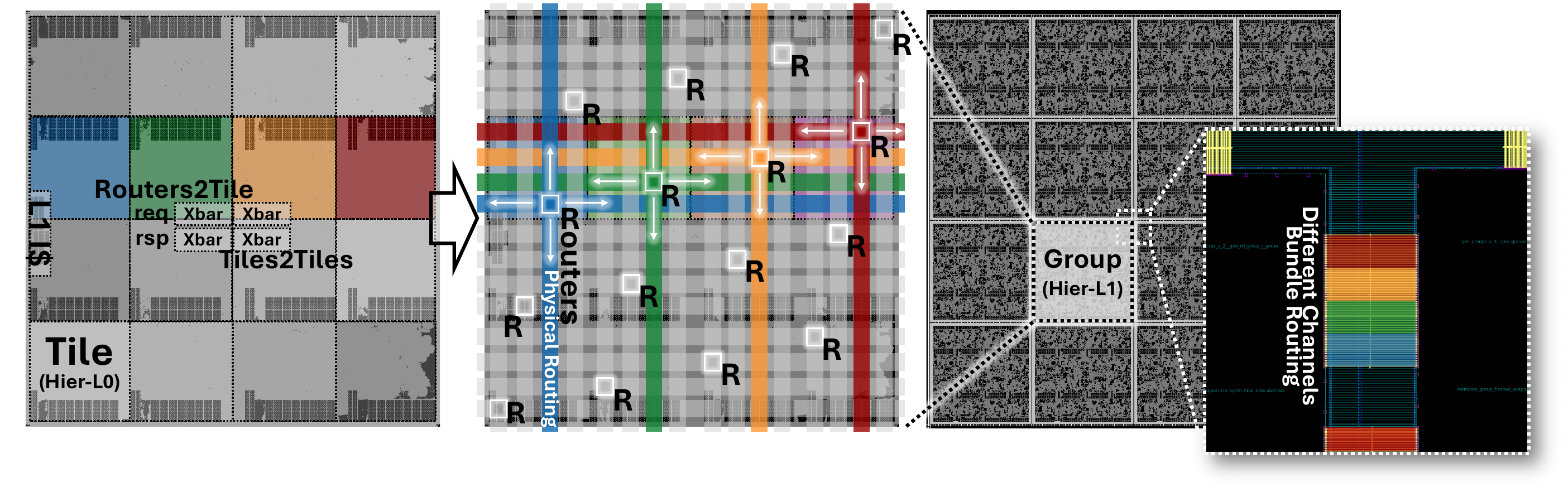}
  \vspace{-3em}
  \caption{Physical layout of the \num{1024}-core shared-L1-memory cluster with annotated TeraNoC interconnect, showing regular and bundled routing patterns.}
  \vspace{-1em}
  \label{fig:physical_view}
\end{figure*}

\subsection{Physical Implementation}
We implement the TeraNoC-based \num{1024}-core, shared-\num{4096}-L1-bank cluster using \emph{GlobalFoundries}' \SI{12}{\nano\meter} \emph{LPPLUS FinFET} technology.
Synthesis and \gls{pnr} are performed with \emph{Synopsys}' \emph{Fusion Compiler} 2023.12, and power consumption is determined using \emph{Synopsys}' PrimeTime 2022.03 under typical operating conditions (TT/\SI{0.80}{\volt}/\SI{25}{\celsius}).

We present the full cluster physical layout in~\Cref{fig:physical_view}.
To fully leverage the available \gls{beol} resources for TeraNoC routing, we flatten the \emph{Tile} within each \emph{Group} to enable routing across over.
The routers-to-\emph{Tile} \glspl{xbar} for incoming request\&response and the intra-\emph{Group} \emph{Tile}-to-\emph{Tile} \glspl{xbar} are centrally placed within each \emph{Group}.
The highlighted \emph{Tiles} illustrate that each router is placed within its corresponding \emph{Tile}, while the router ports from different \emph{Tiles} are interleaved along the \emph{Group} boundary.
Routing channels for each router are constrained per direction to ensure straight, shortest-distance paths, and wires in the same direction are easily bundled to facilitate inter-\emph{Group} routing.

\begin{figure}[htbp]
  \centering
  \includegraphics[width=\linewidth]{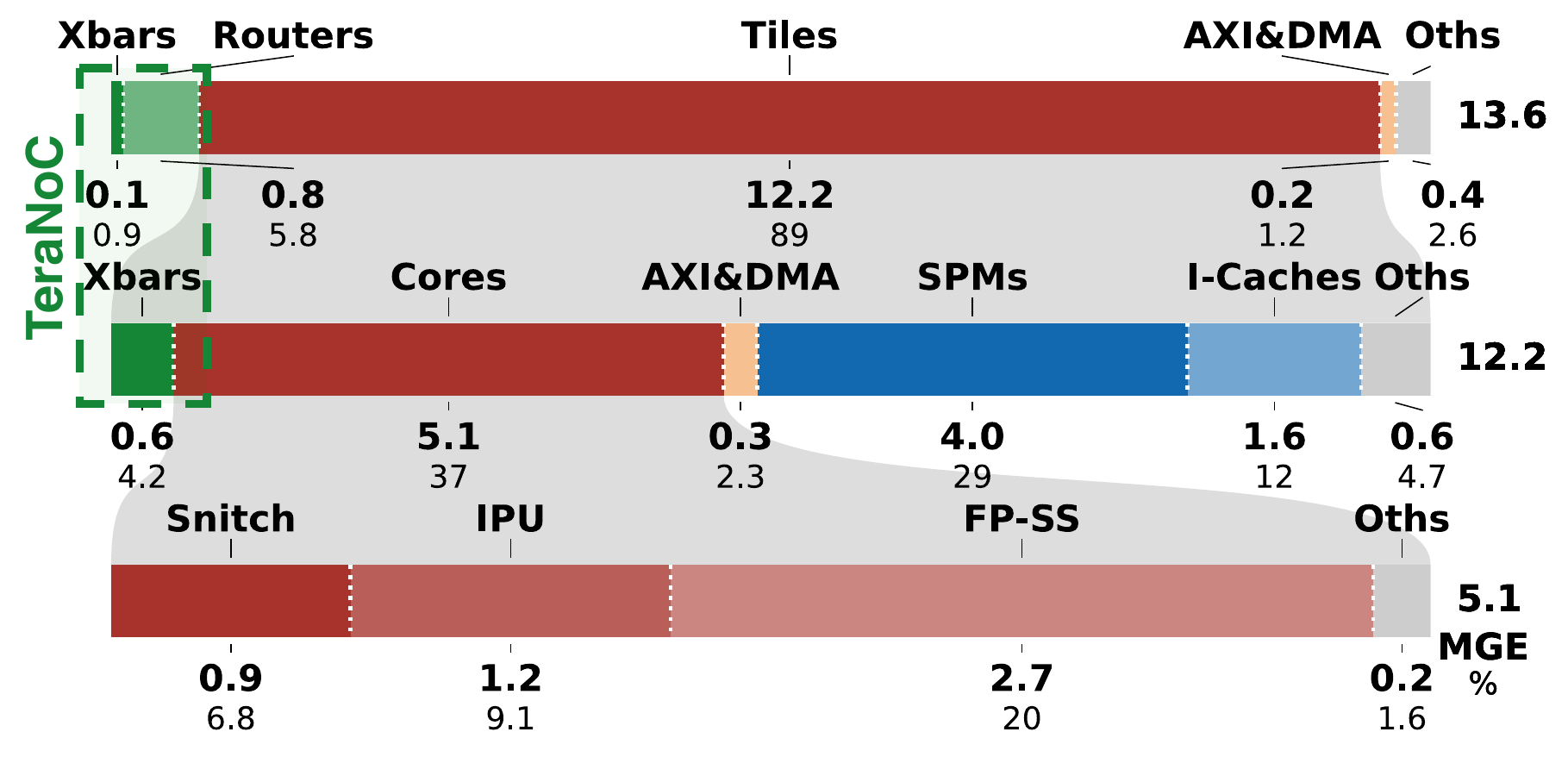}
  \vspace{-2em}
  \caption{Logic area breakdown in gate equivalents for \emph{Group} implementation.}
  \vspace{-1em}
  \label{fig:logic_area}
\end{figure}

\begin{figure}[htbp]
  \centering
  \includegraphics[width=\linewidth]{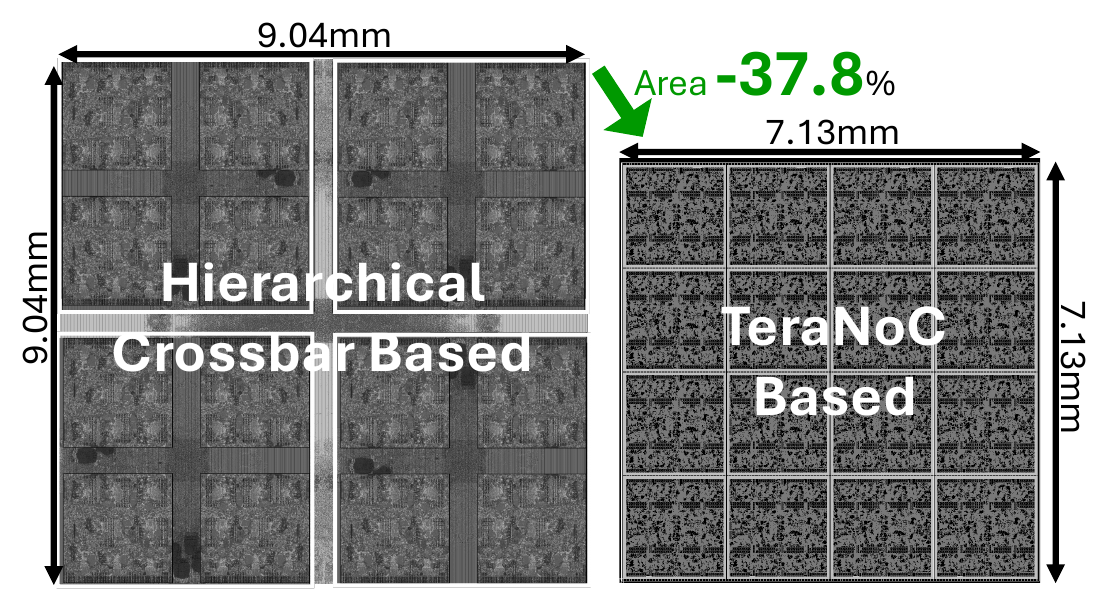}
  \vspace{-2em}
  \caption{The area improvement of TeraNoC-based \num{1024} cores shared-\SI{4}{\mebi\byte}-L1 cluster in \emph{GF12nm} over the hierarchical crossbar-based interconnect solution.}
  \vspace{-1em}
  \label{fig:physical_area}
\end{figure}

\Cref{fig:logic_area} shows the \emph{Group} area breakdown in \gls{ge}.
Most of the logic area is dedicated to computation \glspl{pe} (\SI{37}{\percent}), data \gls{spm} (\SI{29}{\percent}), and \gls{IDol} (\SI{12}{\percent}), while the TeraNoC core-to-L1-bank interconnect accounts for only \SI{10.9}{\percent}.
\Cref{fig:physical_area} compares the physical die area with the hierarchical \gls{xbar}-based interconnect implementation of TeraPool cluster (\emph{GF12nm FinFET}).
The TeraNoC solution significantly reduces the physical routing area, leaving tiny gaps between \emph{Groups} for only global signals routing (clock, scan, reset, control).
Overall, the total cluster area is reduced by \SI{37.8}{\percent}.
With our solution, the cluster achieves a frequency of \SI{936}{\mega\hertz} (TT/\SI{0.80}{\volt}/\SI{25}{\celsius}), representing a \SI{13.3}{\percent} increase compared with the hierarchical-\gls{xbar}-based baseline cluster\footnote{\terapool{} configuration with remote \emph{Group} latency of 9 cycles.} at \SI{850}{\mega\hertz}, as the interconnect is no longer on the critical path.

\subsection{Software Evaluation}
A key advantage of scaling up a shared-memory cluster is its programming friendliness: a large unified-address shared-L1 space simplifies data movement, splitting, and merging compared to multiple loosely coupled private-L1 clusters interconnected by long-latency global interconnects.
Our testbed cluster supports a streamlined \emph{fork-join} programming model~\cite{Zhang_2024} for data-parallel computing C-runtime, enabling efficient transitions between sequential control and parallel computing in \gls{spmd} execution.
We evaluate the TeraNoC-based cluster performance using key data-parallel kernels for embodied \gls{gai} workloads.
The kernels include both \textbf{\emph{local access}}-dominated workloads, where \glspl{pe} primarily fetch data from a portion of the shared memory through low-latency \glspl{xbar}; and \textbf{\emph{global access}}-dominated workloads, where \glspl{pe} fetch distributed data structures across all \glspl{spm} via the 2D-mesh \gls{noc}, as detailed below:
\begin{itemize}[leftmargin=*]
    \item \textbf{\gls{axpy}} is a representative \emph{local access}-dominated kernel, widely used in embodied systems for physics-based control, gradient updates, and residual connections. We parallelize all \glspl{pe} to fetch, compute, and store on their local portion of shared memory via TeraNoC’s lower-hierarchy \glspl{xbar}.
    \item \textbf{\gls{dotp}} computes the scalar product of two vectors, a common pattern in AI-enhanced environment sensing, such as neural activations and attention score computations. It is parallelized similarly to \gls{axpy}, with each core accumulating into a private reduction variable, followed by a final reduction stage that requires global data synchronization through 2D-mesh.
    \item \textbf{\gls{gemv}} is a key operator in transformer model training, dense layer inference, and value function estimation, extending \gls{dotp} to entire layers, with a reduction step required for each matrix row to accumulate partial dot products across \glspl{pe}.
    \item \textbf{\gls{conv2d}} is the dominant operation in convolutional neural networks and early-stage perception networks. The weights are distributed into each \gls{pe}’s local \emph{Tile} for repetitive fast access. \glspl{pe} fetch input matrix mainly from the local and neighboring \emph{Tiles/Groups}, benefiting from the lower latency of the \emph{intra-Group} interconnect.
    \item \textbf{\gls{matmul}} is the most compute-intensive operator in multi-head attention mechanisms. The matrix’s fully interleaved row/column data across all banks makes it an extremely \emph{global access}-dominated kernel. We employ a $4 \times 4$ tiled parallelization to fully utilize the register file and maximize computational intensity. Each \gls{pe} shifts its fetching offsets to reduce potential hierarchical interfaces and bank conflicts.
\end{itemize}

\begin{figure}[htbp]
  \centering
  \includegraphics[width=\linewidth]{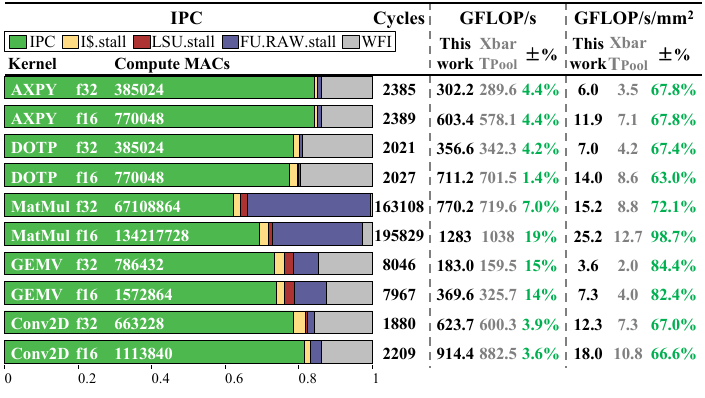}
  \vspace{-2em}
  \caption{Key \gls{gai} kernels' IPC breakdown; performance and area efficiency are compared with the crossbar-based TeraPool cluster (Xbar-TPool).}
  \label{fig:software_performance}
\end{figure}

We benchmark the kernels using the largest input size that fits into the cluster’s L1 \gls{spm}.
The results are shown in~\Cref{fig:software_performance}, which annotates the kernels' \gls{mac} complexity and execution cycles.
The performance is compared with the \gls{xbar}-based cluster baseline under typical operating conditions (TT/\SI{0.80}{\volt}/\SI{25}{\celsius}).
\gls{axpy}, \gls{dotp}, and \gls{gemv} serve as \emph{local-access} dominated kernels to evaluate TeraNoC’s \emph{Intra-Group} \glspl{xbar} performance.
A high \gls{ipc} of up to \num{0.85} is achieved, with a slight loss mainly due to synchronization (\gls{wfi}) at the end of parallel execution, which is slightly higher for \gls{dotp} and \gls{gemv} because of the necessary sum reduction across \glspl{pe}.
\gls{conv2d} utilizes both \glspl{xbar} and the 2D-mesh NoC for \emph{intra-/inter-Group} access and achieves a high \gls{ipc} of \num{0.82}. Thanks to the localized weights and short-distance input matrix fetching from neighboring \emph{Tiles/Groups}, only \SI{1}{\percent} of cycles are attributed to \gls{lsu} stalls.
For \gls{matmul}, a global-access dominated kernel that places extremely high pressure on the 2D-mesh channels for long-distance requests and response, the \gls{ipc} still remains high at \num{0.7} for both single- and half-precision execution.
Compared to hierarchical multi-stage \glspl{xbar}, TeraNoC’s multi-channel 2D-mesh can handle more concurrent requests without conflicts.
The observed \gls{ipc} loss of \gls{matmul} and \gls{gemv} is not attributed to interconnect conflicts (\gls{lsu} stalls), but rather to execution functional units waiting for response data (FU.\gls{raw} stalls) from long-distance remote banks.
In addition, the limited number of registers (\num{32}) in the RISC-V \gls{isa} prevents further scheduling more outstanding requests for latency hiding.

\begin{figure}[htbp]
  \centering
  \includegraphics[width=\linewidth]{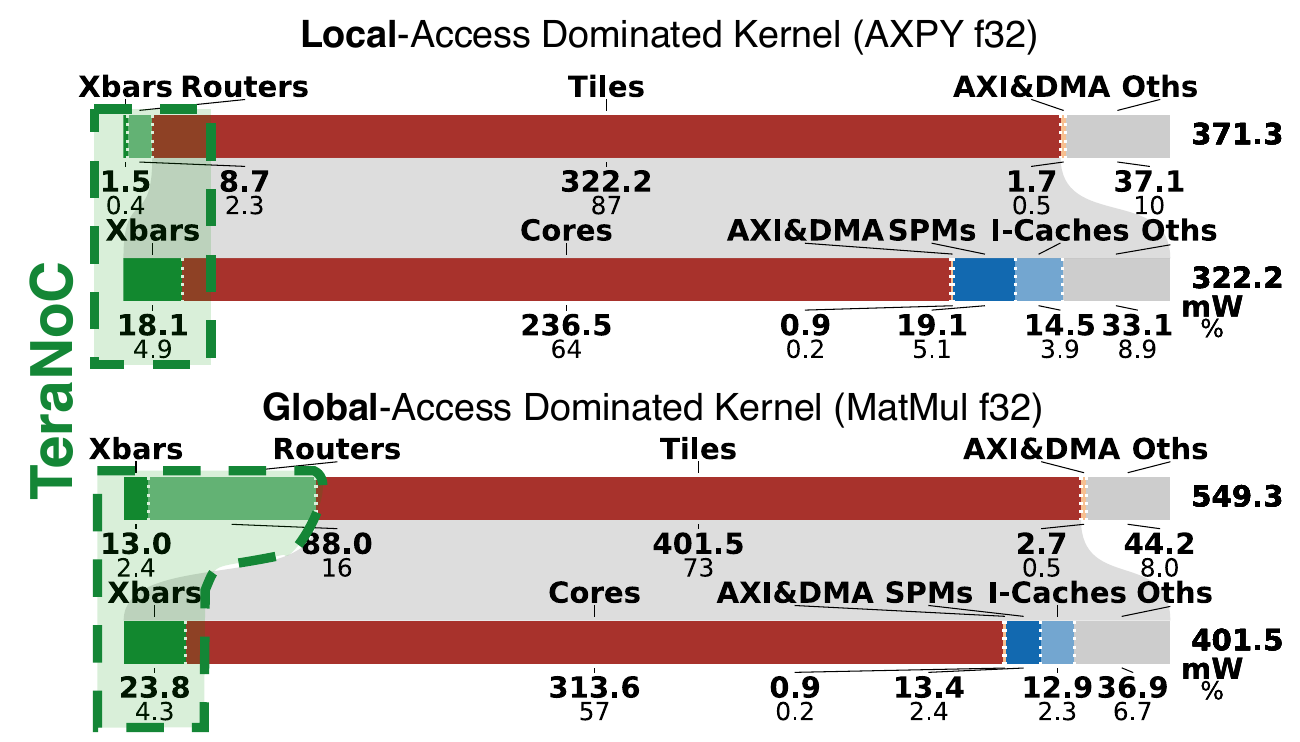}
  \vspace{-2em}
  \caption{Power breakdown (\emph{Group}) for local-\&global-access dominated kernels.}
  \label{fig:kernel_power}
\end{figure}

Across the benchmarked kernels, the TeraNoC-based cluster achieves up to \SI{0.77}{\tera\flop\per\second} in single-precision and \SI{1.3} {\tera\flop\per\second} in half-precision performance, delivering up to a \SI{19}{\percent} improvement in throughput (\si{\giga\flop\per\second}) and a \SI{98.7}{\percent} increase in area efficiency (\si{\giga\flop\per\second\per\square\milli\meter}) compared to the hierarchical \gls{xbar}-based cluster baseline.
\Cref{fig:kernel_power} shows the \emph{Group} power breakdown for both \emph{local-} and \emph{global-access} dominated kernels.
TeraNoC consumes only \SI{7.6}{\percent} of the total cluster power for kernels primarily accessing the \glspl{xbar}.
Even for kernels with \gls{pe}-to-L1 remote traffic patterns that traverse extremely long distances, where heavy traffic loads stress the 2D-mesh channels through multiple routers, TeraNoC consumes only \SI{22.7}{\percent} of the total cluster power, demonstrating a highly efficient scaled-up cluster interconnect.

\section{Conclusion}
\label{sec:conclusion}
In this paper, we presented TeraNoC, a hybrid mesh-crossbar, \SI{32}{\bit} fine-grained multi-channel on-chip interconnect for core-to-L1-bank connections, enabling area-efficient scaling up of shared-memory clusters.
We built a cluster that matches the largest tightly coupled L1 cluster reported in the literature~\cite{Zhang_2024}, comprising \num{1024} cores sharing 4096 \gls{spm} banks.
TeraNoC achieved a peak bandwidth of \SI{3.74}{\tebi\byte\per\second} and a bisection bandwidth of \SI{0.47}{\tebi\byte\per\second}, while occupying only \SI{10.9}{\percent} of the total logic area.
A router-remapper balanced traffic loads across different channels, improving bandwidth utilization by $2.7\times$.
In benchmarked key data-parallel kernels for embodied \gls{gai} workloads, TeraNoC’s high bandwidth maintained high cluster compute utilization, achieving \gls{ipc} up to \num{0.85}, while consuming only \num{7.6}\textendash\SI{22.7}{\percent} of total cluster power.
Compared to the hierarchical multi-stage \gls{xbar}-based cluster baseline, TeraNoC reduced total cluster area by \SI{37.8}{\percent} and increased computing throughput by \SI{19}{\percent}, improving area efficiency (\si{\giga\flop\per\second\per\square\milli\meter}) by up to \SI{98.7}{\percent}.
 
\section*{Acknowledgment}
\ifx\blind\undefined
    \noindent\small{This work has received funding from the Swiss State Secretariat for Education, Research, and Innovation (SERI) under the SwissChips initiative.}
\else
    \textit{Omitted for blind review.} \\ \\
\fi

\Urlmuskip=0mu plus 1mu\relax
\def\UrlBreaks{\do\/\do-}
\bibliographystyle{IEEEtran}
\bibliography{bibliography/bibliography}

\end{document}